\documentstyle[12pt]{article}

\newcommand{\be}{\begin{equation}}
\newcommand{\ee}{\end{equation}}
\newcommand{\bea}{\begin{eqnarray}}
\newcommand{\eea}{\end{eqnarray}}
\newcommand{\bd}{\begin{displaymath}}
\newcommand{\ed}{\end{displaymath}}
\newcommand{\bi}{\begin{itemize}}
\newcommand{\ei}{\end{itemize}}
\newcommand{\bc}{\begin{center}}
\newcommand{\ec}{\end{center}}
\newcommand{\bfl}{\begin{flushleft}}
\newcommand{\efl}{\end{flushleft}}
\newcommand{\bfr}{\begin{flushright}}
\newcommand{\efr}{\end{flushright}}

\def\6{\partial} \def\a{\alpha} \def\b{\beta}

\def\m{\mu} \def\n{\nu}  
 \def\ss{\sigma} \def\t{\tau}

 \def\G{\Gamma}

\def\={\!\!\!&=&\!\!\!}
\def\+{\!\!\!&&\!\!\!+~}
\def\-{\!\!\!&&\!\!\!-~}

\newcommand{\HH}{{\cal H}}

\newcommand{\LL}{{\cal L}}

\newcommand{\OO}{{\cal O}}

\begin{document}

\title{$D$-Branes at Finite Temperature in TFD}	
\author{M. C. B. Abdalla${}^a$\footnote{mabdalla@ift.unesp.br}, A. L. 
Gadelha${}^a$\footnote{gadelha@ift.unesp.br},
I. V. Vancea${}^b$\footnote{ion@dfm.ffclrp.usp.br}}
\maketitle

\begin{center}
{\small ${}^a$ Instituto de F\'{\i}sica Te\'{o}rica, Universidade 
Estadual Paulista\\
Rua Pamplona 145, 01405-900, S\~{a}o Paulo, SP, Brasil\\
and\\
${}^b$Departamento de F\'{\i}sica e Matem\'{a}tica,\\ Faculdade de 
Filosofia, Ci\^{e}ncias e Letras de Ribeir\~{a}o Preto
USP,\\ Av. Bandeirantes 3900, Ribeir\~{a}o Preto 14040-901, SP, Brasil}
\end{center}

\begin{abstract}
We review the construction of the $D$-branes at finite temperature as 
boundary states in the Fock space of thermal perturbative closed string. 
This is a talk presented by I. V. V. at Common Trends in Cosmology and 
Particle Physics June 2003, Balatonf\"{u}red, Hungary.
\end{abstract}

\newpage


Since their discovery in 1995 \cite{pol}, $D$-branes have played a 
major role in the development of the string theory and the models based on 
it. They have given a first basis to unify the five superstring 
theories, offered powerfull tools to understand the nonperturbative limit of 
string theory, made possible the connection between field theory and 
gravity and have been used to address some fundamental problems as the 
black hole entropy and hierarchy problem, to mention just some of the 
applications of the $D$-branes. Also, new mathematical tools have had to be 
employed and new mathematical problems emerged in the investigation of 
the properties of the $D$-branes. By its richness in concepts and 
applications, $D$-brane theory represents one of the most fruitful research 
direction in the modern mathematical physics. 

One of the most interesting problems concerning the $D$-branes is the 
development of a microscopic description of their thermodynamical 
properties. In the low energy limit of string theory, i. e. when the string 
tension is infinite and strings behave as particles, the $D$-branes are 
described by solutions of supergravity. Thus, the thermodynamics of the 
$D$-branes can be computed using the Masubara formalism. However, one 
can describe the $D$-branes as boundary states in the Fock space of the 
closed string theory by interpreting the one-loop diagram in open 
string channel as the tree-level diagram in the closed string channel or by 
using the T-duality. Then, one can naturaly ask: What is the state 
corresponding to the $D$-brane if the string is thermal? Answering this 
question gives us the description of the $D$-brane at finite temperature 
as a quantum state in the Fock space of the thermal closed string 
\cite{ivv1,ivv2,ivv3,ivv4,ivv5}. We note, however, that in the perturbative 
limit of string theory the $D$-branes can be viewed as (physical) 
submanifolds of the space time where the open strings can end. In this case, 
one can compute the free energy of the open strings in the presence of 
the $D$-branes and obtain the self-energy of the $D$-brane through the 
finite temperature dualities \cite{vm}. However, this method does not 
give us the representation of the $D$-brane as a thermal state.

In order to answer the above question, we have to be able to manipulate 
the transformation from zero temperature to finite temperature at the 
Fock space level. A simpler way to do that is to use the Thermo Field 
Dynamics (TFD) formalism \cite{umezbook} in which the statistical average 
of an observable ${\cal O}$ is expressed as a "thermal vacuum" 
expectation value 
\begin{equation}
\langle {\cal O} \rangle = \langle\langle 0(\beta_T )|{\cal O} | 
0(\beta_T ) \rangle\rangle .
\label{average} 
\end{equation}  
The "thermal vacuum" state $| 0(\beta_T ) \rangle\rangle $ belongs to 
the Fock space that is a direct product between the original Fock space 
and an independent identical copy of it. 
This second copy of the original string is denoted by a tilde and does 
not represent a physical system. The $\beta_T$ states for the inverse of 
the temperature in units such that the Botzmann constant is one. The 
temperature is implemented in the Fock space of the doubled system by a 
Bogoliubov transformation $\Gamma(\beta_T )$ that mixes operators from 
the two copies
\bea
| 0(\beta_T ) \rangle\rangle &=& e^{-i\G(\beta_T )}| 0 \rangle\rangle 
\label{thvac}\\
\cal{O}(\beta_T ) &= &e^{-i\Gamma(\beta )} \OO e^{i\Gamma (\beta_T )},
\label{thobs}
\eea
where $\left.\left| 0 \right\rangle\right\rangle = \left| 0 
\right\rangle\widetilde{\left| 0\right\rangle}$ is the product vacuum state 
between the two vacua of the string and the tilde string.
In general, the possible Bogoliubov operators form a group of 
transformations, which for bosons is $SU(1,1)$, while for fermions is $SO(2)$. 
For internal consistency of the theory, the thermal vacuum should be 
invariant under the tilde operation that transforms the two copies of the 
Fock space one into another. One can select an unique Bogolibov 
operator that guarantees the tilde invariance and generates an unitary 
transformation \cite{umezbook}. This represents our choice. 

Our purpose now is to apply the TFD method to string theory in order to 
find the state corresponding to the $D$-brane at finite temperature. 
The Fock space of the closed string is a direct product between the Fock 
spaces of the left- and right-moving oscillators as well as the Hilbert 
space of the center of mass degrees of freedom. If the string is 
supersymmetric on the world-sheet, one should tensor this Fock space with two 
more spaces, corresponding to the left- and right-moving 
two-dimensional Majorana spinors. Since all modes are independent of each other, one 
can apply the TFD construction separately to each Fock space. If we 
consider for simplicity just the bosonic string, the Bogoliubov 
transformation that takes the zero temperature Fock space 
\be
\hat{\HH} = \HH\otimes \tilde{\HH},
\label{totalfockzero}
\ee
where tilde denotes the second copy of the bosonic string Fock space, 
to finite temperature, has the following form
\be
\G = \G^{\a} + \G^{\b} 
=- i \sum_{n=1}^{\infty}\theta_{n}(\b_T)
(A_{n}\cdot{\tilde{A}}_{n} - 
A^{\dagger}_{n}\cdot{\tilde{A}}^{\dagger}_{n} 
+ B_{n}\cdot{\tilde{B}}_{n} - 
B^{\dagger}_{n}\cdot{\tilde{B}}^{\dagger}_{n}).
\label{gop}
\ee
Here, the superscripts $\a$ and $\b$ denote the left- and right-moving 
modes.
The operators $A^{\m}_n$ and $A^{\m \dagger}_n$ annihilate and create 
string left-moving excitations of frequency $n$ and in the space-time 
direction $\m = 1,2,\ldots ,24$. All the directions are space-like since 
we work in the light-cone gauge in order to elliminate the unphysical 
degrees of freedom. The same considerations apply to the right-moving 
modes $B$. The Bogoliubov operator depends on the temperature through the 
paramenters $\theta$ which, for bosons, are given by the following 
equation
\be
\cosh \theta_n(\b_T) = (1-e^{-\b_T n})^{-\frac{1}{2}}.
\label{theta}
\ee

The Bogoliubov operator (\ref{gop}) acts on $X^{\m}(\t , \ss )$ which 
represents the most general solution of the closed string equations of 
motion. ($X^{\m}(\t , \ss )$ is a linear combination between a linear 
term in the world-sheet time-like parameter $\t$ and independent left- 
and right-moving oscillations \cite{jpbook}). The resulting operator, 
denoted by $X^{\m}(\t , \ss )(\beta_T)$
is still a solution of the string equations of motion with the creation 
and annihilation operators now at finite temperature given by 
(\ref{thobs}). The coordinates and momenta of the center of mass are inert under 
the Bogoliubov transformation. Since we work with two copies of the 
originals string, we have to apply (\ref{gop}) to the tilde string as well 
and we obtain the second copy of the string $\tilde{X}^{\m}(\t , \ss 
)(\beta_T)$ at finite temperature. The creation and annihilation 
operators $A^{\m}_{n}(\beta_T)$,  ${\tilde{A}}^{\m}_{n}(\beta_T)$, $A^{\m 
\dagger}_{n}(\beta_T)$, ${\tilde{A}}^{\m \dagger}_{n}(\beta_T)$, 
$B^{\m}_{n}(\beta_T)$, ${\tilde{B}}^{\m}_{n}(\beta_T)$,$ B^{\m 
\dagger}_{n}(\beta_T)$ and ${\tilde{B}}^{\m \dagger}_{n}(\beta_T)$ act on the thermal 
vacuum given by (\ref{thvac}). All the oscillator properties of each string 
mode are preserved at finite temperature. 

Let us take a look at the conformal symmetry. By using the Bogoliubov 
transformations (\ref{thvac}) and (\ref{thobs}) we have contructed 
solutions of the string equations of motion at finite temperature. These 
solutions are obtained basically by mapping the creation and annihilation 
operators for each string mode at finite temperature. Then the 
generators of the conformal transformations, the Virasoro operators which can 
be expressed in terms of creation and annihilation operators are also 
mapped at finite temperature. Since the Bogoliubov transformation is 
unitary, the Virasoro operators at finite temperature $L_n(\beta_T)$ 
satisfy the Virasoro algebra. However, if one computes physical quantities, 
one has to employ observables at zero temperature according to the 
assumption (\ref{average}). It is a simple exercise to show that the thermal 
vacuum does not vanish under the action of the Virasoro operators at 
zero temperature $L_n$ and thus the thermal vacuum breaks the zero 
temperature Virasoro algebra. Similarly, the true vacuum (at zero 
temperature) breaks the Virasoro algebra at finite temperature. Breaking the 
symmetries at finite temperature by the zero temperature operators is a 
general property of the thermal systems and it is not specific to strings, 
as we have used the general arguments of TFD do derive it. This remarks 
were also made in \cite{susytfd} in the case of the supersymmetric 
oscillator. 

Let us turn now to the discussion of the $D$-branes. If one imposes the 
Neumann boundary conditions along the directions $a = 1, 2, \ldots , p$ 
and the Dirichlet boundary conditions along $i = p+1, \ldots , 24$ in 
the open string theory, one obtaines a hypersurface that exchanges 
momenta with the string and defines the $Dp$-brane. By employing the 
T-duality, the $Dp$-brane is mapped into the closed string channel. This 
mapping changes the Dirichlet and the Neumann boundary conditions among 
themselves. It also suggests the following interpretation of the two types 
of strings in $D$-brane theory: the open string channel describes the 
degrees of freedom of the $D$-brane, while the closed strings describe 
the interactions of two or more $D$-branes with one another.  
If we denote by $X^a (\t ,\ss )$ and $X^i (\t , \ss )$ the closed 
string coordinate operators along the Dirichlet and the Neumann directions, 
respectively, the equations defining a $Dp$-brane are the following 
ones:
\bea
\6_{\t}X^{a}|_{\t = 0}\left| B \right\rangle &=& 0 \nonumber\\
X^{i}|_{\t = 0} \left| B \right\rangle &=& x^i ,
\label{defDbr}
\eea
where $x^i$ denote the coordinates of the hyperplane in the transverse 
space. If one Fourier expands the string coordinate operators, the 
equations (\ref{defDbr}) become conditions on the Fock space of the closed 
string. The solution to this set of equations is given by the following 
relation:
\be
\left| B\right\rangle =N_{p}\delta ^
{\left(d_{\perp }\right) }\left(q-x\right)
e^{ -\sum\limits_{n=1}^{\infty }A
_{n}^{\dagger\mu }S_{\mu \nu }B_{n}^{\dagger\nu }}\left| 
0\right\rangle,
\label{soltempzero}
\ee
where $q^{\m}$ are the operators corresponding to the center of mass of 
the closed string, $S_{\m \n }$ is the diagonal matrix $S_{\m \n} = 
\left( \eta_{ab},-\delta_{ij} \right)$ and $\left| 0 \right\rangle$ is the 
translationally invariant closed string vacuum. The same boundary 
conditions should be imposed on the copy of the Fock space and they generate 
a $Dp$-brane solution in the tilde operators. 

The counterpart of the $Dp$-brane  at finite temperature should be 
obtained by applying the same idea that has led to the coherent state 
(\ref{soltempzero}) at zero temperature. Indeed, if one varies continuously 
the temperature $T$ from zero to a small but finite value, the 
hypersurface that defines the $Dp$-brane at zero temperature should not change. 
Therefore, one should be able to use the same geometrical condition, i. 
e. the same Dirichlet and Neumann boundary conditions, to write down 
the equations that define the thermal coherent state in the Fock space of 
the thermal string and to interpret the solution to these equations as 
a thermal $Dp$-brane. As we have already seen, the fields $X^{\m}(\t, 
\ss)(\b_T)$ and $\tilde{X}^{\m}(\t , \ss )(\b_T)$ satisfy the string 
equations of motion that can be obtained from the variational principle 
applied to the Lagrangian density $\hat{\LL} = \LL-\tilde{\LL}$ of the 
two copies of the closed string. If we imposed the Dirichlet and the 
Neumann boundary conditions to the action $\hat{S}$, we obtain two copies 
of the relations (\ref{defDbr}) but with the string operators at finite 
temperature instead of those at zero temperature:
\bea
\6_{\t}X^{a}|_{\t = 0}(\b_T)\left.\left| B(\b_T) \right\rangle 
\right\rangle&=& 
\6_{\t}\tilde{X}^{a}|_{\t = 0}(\b_T)\left.\left| B(\b_T) 
\right\rangle\right\rangle
=0 \nonumber\\
X^{i}|_{\t = 0}(\beta_T) \left.\left| B(\b_T) 
\right\rangle\right\rangle &=&
\tilde{X}^{i}|_{\t = 0}(\beta_T) \left.\left| B(\b_T) 
\right\rangle\right\rangle = x^i,
\label{defDbrT}
\eea 
where the coordinates of the centers of mass of the string and the 
tilde string are the same. The $Dp$-brane state $\left.\left| B(\b_T) 
\right\rangle\right\rangle$ has now contributions from the two sectors of 
the Fock space $\hat{H}$. In order to find the solution of the equations 
(\ref{defDbrT}) we note that the algebra of the string oscillators at 
finite temperature is preserved by the Bogoliubov transformations, which 
means that the temperature dependent creation and annihilation 
operators and tilde operators are independent. The thermal $Dp$-brane solution 
has the following form
\be
\left.\left| B(\b_T) \right\rangle\right\rangle = 
N_p^2\delta^{2\left(d_{\perp }\right) }\left(q-x\right)
e^{-\sum\limits_{n=1}^{\infty }\left[A
_{n}^{\dagger\mu }(\b_T)+\tilde{A}
_{n}^{\dagger\mu }(\b_T)\right]
S_{\mu \nu }\left[B_{n}^{\dagger\nu }(\b_T)
 + \tilde{B}_{n}^{\dagger\nu }(\b_T)\right]}
\left.\left| 0 (\b_T)\right\rangle\right\rangle,
\label{thermDp}
\ee 
where $\delta^{2\left(d_{\perp }\right) 
}\left(q-x\right)=\delta^{\left(d_{\perp }\right) }\left(q-x\right)
\delta^{\left(d_{\perp }\right) }\left( \tilde{q}-x\right)$. One can 
easily  check that the same solution can be obtained by mapping the 
product of $Dp$-brane state and tilde state at zero temperature to $ T \neq 
0$. Also, we can obtain partial $Dp$-brane-like solutions in either 
thermal string or tilde string sectors by solving the constraints 
(\ref{defDbrT}) in the respective variables.

As an immediate application of the TFD formalism in solving the thermal 
$Dp$-branes is the calculation of their "internal" entropy as follows. 
In the TFD approach, one can define the entropy operator of an infinite 
set of independent oscillators. If we consider the closed string as 
beeing the set of independent oscillation modes, its entropy in $k_B$ 
units is given by the following relation
\begin{eqnarray}
K &=&\sum\limits_{\mu }\sum\limits_{n}\left[ \left( A_{n}^{\mu \dagger
}A_{n}^{\mu }+B_{n}^{\mu \dagger }B_{n}^{\mu }\right) \log \sinh 
^{2}\theta
_{n}\right. +  \nonumber \\
&&\left. -\left( A_{n}^{\mu }A_{n}^{\mu \dagger }+B_{n}^{\mu 
}B_{n}^{\mu
\dagger }\right) \log \cosh ^{2}\theta _{n}\right]. 
\label{entroop1}
\end{eqnarray}
This operator commutes with the Bogoliubov generator. A similar entropy 
can be defined for the tilde string, but it does not describes a 
physical quantity according to the TFD principles.
At the boundary of the world-sheet, the string modes generate a 
coherent state which represents the $Dp$-brane. Thus, the entropy of the 
$Dp$-brane is given by the expectation value of the entropy operator 
(\ref{entroop1}) in the state (\ref{thermDp})
\begin{eqnarray}
K_{Dp} = \left\langle \! \left\langle 
B\left( \beta _{T}\right) \right| \right.
K\left. \left| B\left( \beta _{T}\right) \right\rangle \! \right\rangle
&=&48\sum_{m=1}^{\infty }\left[\log \sinh ^{2}\theta _{m}
-{\sinh}^{2}\theta_m \log \tanh^{2}\theta _{m}\right] \nonumber \\
&&+2\prod_{m=1}^{\infty }\prod_{\mu=1}^{24}
\prod_{\nu =1}^{24}\sum_{k=0}^{\infty }{\cosh}^{2}\theta_m
\frac{\left( -\right)^{2k+2}S_{\mu \nu }^{2k+2}}{k!\left( k+1\right) 
!}.
\label{braneentropy}
\end{eqnarray}
The contribution of the oscillators to $K_{Dp}$ diverges
in the $T\rightarrow 0$ limit and behaves as $\log(-1)$ as $ 
T\rightarrow \infty$. This might be an indication that
the notion of temperature breaks down for arbitrary large temperature 
due to 
the similar phenomenon which occurs in string theory at Hagedorn 
temperature.
 
In conclusion, we have shown that there are thermal states in the Fock 
space of the thermal closed bosonic string corresponding to the 
$Dp$-branes. These state represent the natural generalization of zero 
temperature coherent states. Although divergent, an entropy can be calculated 
for closed strings in this state. In the same way, one can introduce 
world-sheet fermions into the formalism and construct thermal states 
corresponding to super $D$-branes. However, we expect that the supersymmetry 
be broken at some stage at least in the case of the space-time 
supersymmetry. We hope to report on this topic in some future work \cite{ivv6}. 

{\bf Acknowledgements}
I. V. V. acknowledges to the organizers of the School and Workshop on 
Common Trends in Cosmology and Particle Physics 2003 held in 
Balatonf\"{u}red, Hungary, for their kind hospitality. A. L. G. was supported by a 
FAPESP postdoc fellowship. I. V. V. was supported by the FAPESP Grant 
02/05327-3.


\begin{thebibliography}{99}
\bibitem{pol}
J.~Polchinski,
Phys.\ Rev.\ Lett.\  {\bf 75}, 4724 (1995)
[arXiv:hep-th/9510017].
\bibitem{vm}
M.~A.~Vazquez-Mozo,
Phys.\ Lett.\ B {\bf 388}, 494 (1996)
[arXiv:hep-th/9607052].
\bibitem{ivv1}
I.~V.~Vancea,
Phys.\ Lett.\ B {\bf 487}, 175 (2000)
[arXiv:hep-th/0006228].
\bibitem{ivv2}
M.~C.~Abdalla, A.~L.~Gadelha and I.~V.~Vancea,
Phys.\ Rev.\ D {\bf 64}, 086005 (2001)
[arXiv:hep-th/0104068].
\bibitem{ivv3}
M.~C.~Abdalla, E.~L.~Graca and I.~V.~Vancea,
Phys.\ Lett.\ B {\bf 536}, 114 (2002)
[arXiv:hep-th/0201243].
\bibitem{ivv4}
M.~C.~Abdalla, A.~L.~Gadelha and I.~V.~Vancea,
Phys.\ Rev.\ D {\bf 66}, 065005 (2002)
[arXiv:hep-th/0203222].
\bibitem{ivv5}
M.~C.~Abdalla, A.~L.~Gadelha and I.~V.~Vancea,
Int.\ J.\ Mod.\ Phys.\ A {\bf 18}, 2109 (2003),
hep-th/0301249
\bibitem{jpbook} J. Polchinski, 
{\em String Theory}, (Cambridge Monographs on Mathematical Physics, 
1998).
\bibitem{susytfd}
R.~Parthasarathy and R.~Sridhar,
Phys.\ Lett.\ A {\bf 279}, 17 (2001)
[arXiv:cond-mat/0006315].
\bibitem{umezbook}
H.Umezawa, H.Matsumoto and M.Tachiki, 
{\em Thermo Field Dynamics and Condensed States}, 
(North Holland, Amsterdam, 1982) 
\bibitem{ivv6} M. C. B. Abdalla, A. L. Gadelha, I. V. Vancea,
{\em Thermal D-Branes in Supersymmetric String Theory}, in preparation
\end{thebibliography}
\end{document}